\documentclass[cits]{PoS}
\usepackage{amssymb}   
\usepackage{amsmath}
\usepackage{subfigure}
\usepackage{bm}

\newcommand {\Bd} {\ensuremath{B^0_d}}
\newcommand {\Bs} {\ensuremath{B^0_s}}
\newcommand {\Bq} {\ensuremath{B^0_q}}

\newcommand {\barBs} {\ensuremath{\bar{B}^0_s}}
\newcommand {\barBq} {\ensuremath{\bar{B}^0_q}}
\newcommand {\asld} {\ensuremath{a^d_{\mathrm{sl}}}}
\newcommand {\asls} {\ensuremath{a^s_{\mathrm{sl}}}}
\newcommand {\aslq} {\ensuremath{a^q_{\mathrm{sl}}}}
\newcommand {\aslb} {\ensuremath{A^b_{\mathrm{sl}}}}

\title{Charge Asymmetries in Semileptonic B Decays}

\ShortTitle{Charge Asymmetries in Semileptonic B Decays}

\author{\speaker{Iain Bertram}%
        \thanks{representing the D0 Collaboration.}\\
       Lancaster University\\
       E-mail: \email{i.bertram@lancaster.ac.uk}}

\abstract{I present measurements made by the D0 collaboration 
 of the time-integrated flavor-specific semileptonic charge asymmetry in  the decays
of the  \Bs\ and \Bd\ mesons that have undergone flavor mixing, \asls\ and \asld , using  10.4 fb$^{-1}$ of proton-antiproton collisions collected by the D0 detector during Run II at the Fermilab
Tevatron Collider. The results are
$\asls = \left[ \rm{-1.12} \pm  0.74\thinspace (\text{stat}) \pm 0.17 \thinspace (\text{syst}) \right]\%$ andf 
$a^d_{\text{sl}}  =  [0.68 \pm 0.45 \text{ (stat.)} \pm 0.14 \text{ (syst.)}]\%$ which are the two most precise published results and are in agreement with  standard model predictions.  
These results are combined with the like-sign dimuon charge asymmetry, \aslb , and  recent preliminary  results from the LHCb and BaBar experiments to obtain new world averages.
}

\FullConference{14th International Conference on B-Physics at Hadron Machines\\
		 April 8-12, 2013\\
		 Bologna, Italy}

\begin{document}

\section{Introduction}
CP violation has been observed in the decay and mixing 
of neutral mesons containing strange, charm and bottom quarks. Currently all measurements of CP violation, 
either in decay, mixing or in the interference between the two, have been consistent with the presence of a single phase in the CKM matrix. 
An observation of  anomalously large CP violation in \Bs\ oscillations can indicate the existence of physics beyond the 
standard model (SM)~\cite{smprediction}. 
Measurements of the like-sign dimuon asymmetry by the D0 Collaboration~\cite{dimuon1, dimuon2} show  evidence 
of anomalously large CP-violating 
effects  using data corresponding to 9~fb$^{-1}$ of integrated luminosity.
Assuming that this asymmetry originates from mixed  neutral $B$ mesons, 
the measured value is $\aslb = C_d \asld + C_s \asls = \left[-0.787 \pm 0.172 \, (\rm{stat.}) \pm 0.021 \, (\rm{syst.}) \right]\%$,
where $a^{s(d)}_{sl}$ is the  time-integrated  
flavor-specific semileptonic charge asymmetry in \Bs (\Bd)  decays  that have undergone flavor mixing 
and $C_{d(s)}$ is the fraction of \Bd (\Bs ) events.  The  value of \asls\ and \asld\ are  extracted from this 
\noindent measurement and found to be
$\asls = (-1.81 \pm 1.06)\%$ and $\asld = (-0.12 \pm 0.52)\%$~\cite{dimuon2}. 
This Note presents a short summary of the  measurements of \asls\ \cite{d0asls} and \asld\ \cite{d0adsl} using the full Tevatron data sample with
an integrated luminosity of  10.4~fb$^{-1}$.

The asymmetry \aslq\ is defined as
\begin{equation}
\aslq = 
\frac
{\Gamma \left(\barBq \rightarrow \Bq \rightarrow \ell^+ \nu X \right) - \Gamma \left(\Bq \rightarrow \barBq \rightarrow \ell^- \bar{\nu} \bar{X} \right)}
{\Gamma \left(\barBq \rightarrow \Bq \rightarrow \ell^+ \nu X \right) + \Gamma \left(\Bq \rightarrow \barBq \rightarrow \ell^- \bar{\nu} \bar{X} \right)},
\end{equation}
where  in these analyses $\ell=\mu$. The flavor of the \Bq\ meson at the time of decay is identified using the charge of the associated muon, and these analyses do not make use of 
initial-state tagging. We assume  there is no production asymmetry between \Bq\ and \barBq\ mesons, 
that there is no direct CP violation in the decay of charm mesons to the indicated states or in the 
semileptonic decay of $B^0_q$ mesons, and that any CP violation in 
\Bq\ mesons  only occurs in mixing.   We also assume that any direct CP violation in the decay of $b$ baryons and charged $B$ mesons is negligible.

\section{The semi-leptonic charge asymmetry in \Bs\ mesons}

The measurement of the semi-leptonic charge asymmetry in \Bs\ mesons  is measured using the decay $\Bs (\barBs) \rightarrow  D_{s}^\mp \mu^\pm X$ decays, with $D_s^{\mp} \rightarrow  \phi \pi^\mp$ and $\phi \rightarrow K^+ K^-$. The measurement is performed using the raw asymmetry
 \begin{equation}
 \label{raw}
A = \frac{ N_{\mu^+D_s^{-}} -  N_{\mu^-D_s^{+}}}{ N_{\mu^+D_s^{-}} + N_{\mu^-D_s^{+}}},
\end{equation}
where  $N_{\mu^+D_s^{-}}$ ($N_{\mu^-D_s^{+}}$) is the number of reconstructed $\Bs \rightarrow  \mu^+ D_s^{-} X$ 
($\barBs \rightarrow  \mu^- D_s^{+} X$) decays.  The time-integrated  
flavor-specific semileptonic charge asymmetry in \Bs\ decays which have undergone flavor mixing, \asls ,  is then given by
 \begin{equation}
\asls \cdot {F_{\Bs}^{\rm{osc}}} = A - A_{\mu} -A_{\rm{track}} - A_{KK} ,
\label{Eq:asls}
\end{equation}
where  $A_{\mu}$ is the reconstruction asymmetry between positive and negatively charged 
muons in the detector~\cite{d0det},  $A_{\rm{track}}$
is the   asymmetry between positive and negative tracks, 
$A_{KK}$ is the residual kaon asymmetry from the decay of the $\phi$ meson, and $F_{\Bs}^{\rm{osc}}$ is the fraction of  $D_s^- \rightarrow  \phi \pi^-$ 
decays that originate from the decay of a \Bs\ meson after a $\barBs \rightarrow \Bs$ oscillation. 
The $F_{\Bs}^{\rm{osc}}$ factor corrects the measured asymmetry for the fraction of events in which the \Bs\ meson is mixed under the assumptions outlined earlier that no other physics asymmetries are present in the other $b$-hadron backgrounds.  The fraction of mixed events integrated over time is extracted using Monte Carlo (MC) simulations.

\begin{figure}[!h]
\centering
\begin{minipage}[t]{.48\textwidth}
\includegraphics[width=0.95\textwidth]{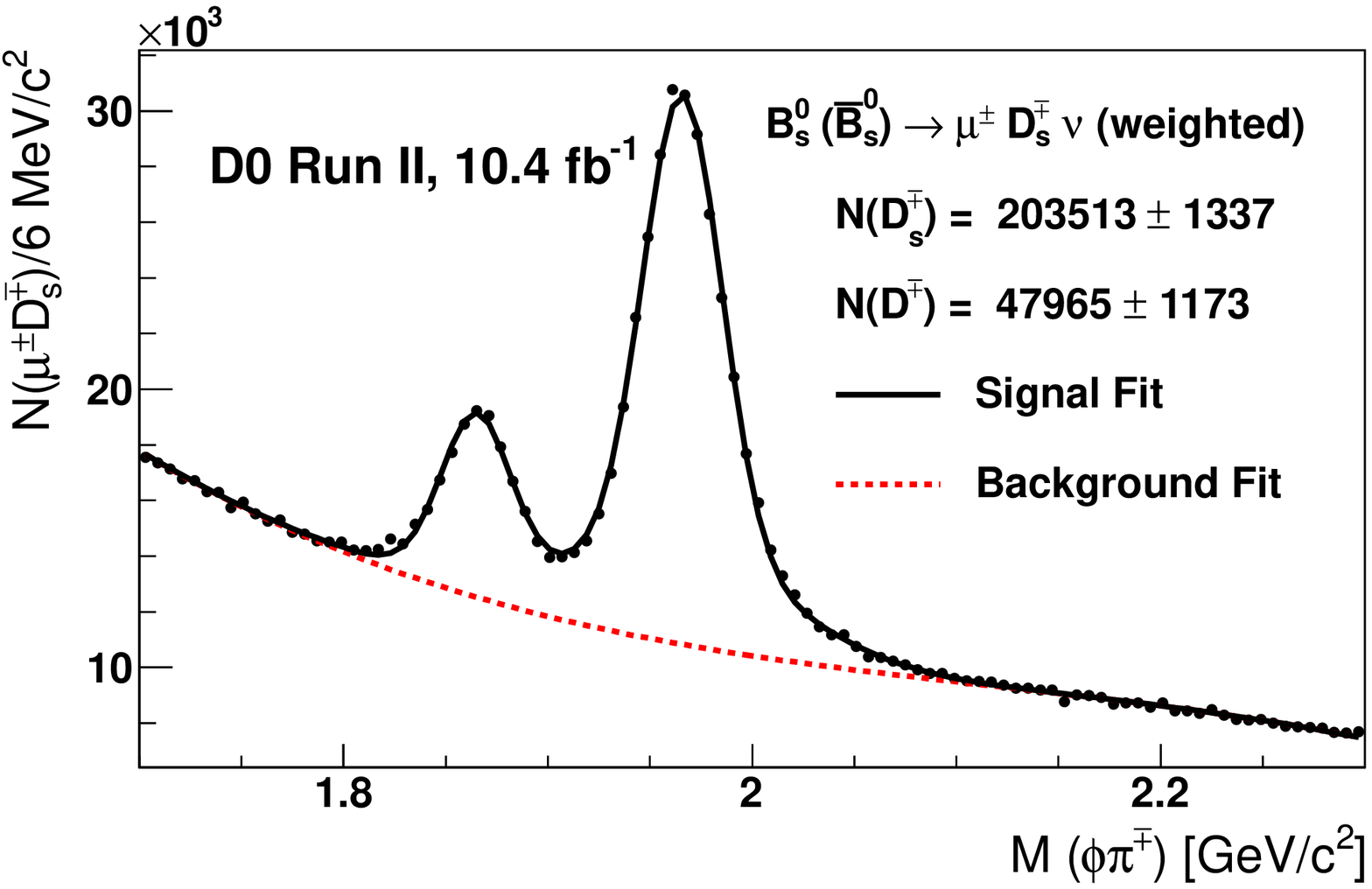}
\caption{\label{Fig:WeightedDSCandidates} 
The weighted $K^+K^-\pi^\mp$ invariant mass distribution for
 the $\mu^{\pm}\phi\pi^\mp$  sample with the  solid line representing the  signal fit and the dashed  line showing the background fit. 
 The lower mass peak is due to  the decay $D^{\mp} \rightarrow \phi \pi^\mp$ and the second peak is due to the $D_s^{\mp}$ meson decay. Note the zero-suppression on the vertical axis.
}
\end{minipage}
\hspace{0.02\textwidth}
\vspace{0pt}
\begin{minipage}[t]{.48\textwidth}
\includegraphics[width=0.95\textwidth]{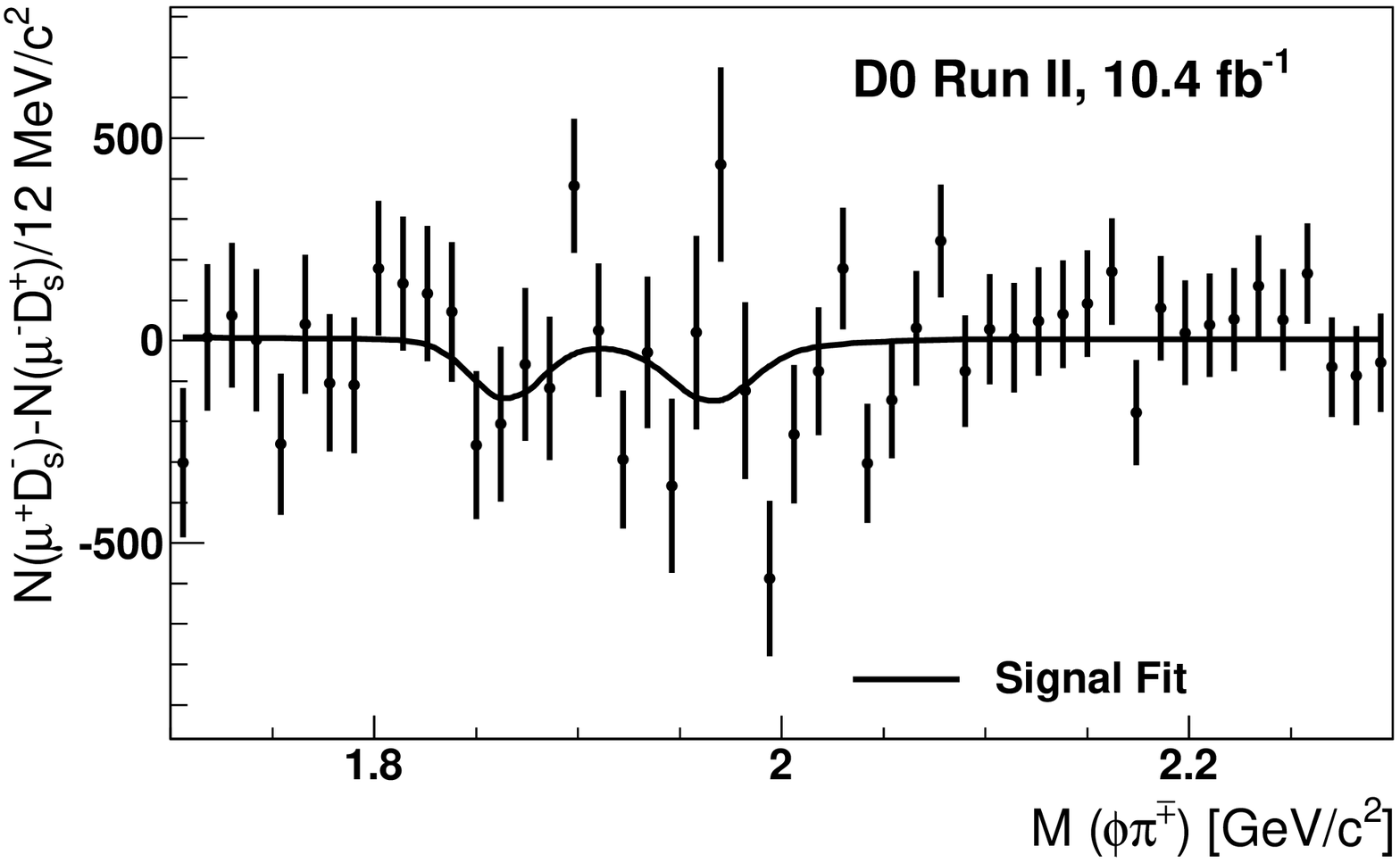}
\caption{\label{Fig:DifferencePlot} 
The fit to the difference distribution for the data with the solid line representing the fit (for clarity the data has been rebinned). 
}
\end{minipage}
\end{figure}

The raw asymmetry  (Eq.~\ref{raw}) is extracted by fitting the $M(\phi\pi^\mp)$ distribution of
the $D_s^\mp$ candidates using a $\chi^2$ minimization. 
The fit is performed simultaneously, using the same models, on the sum (Fig.~\ref{Fig:WeightedDSCandidates}) and
the difference (Fig.~\ref{Fig:DifferencePlot}) of the $M(\phi\pi^-)$ distribution associated with a positively charged muon and $M(\phi\pi^+)$ distribution associated with a negatively charged muon.

The  raw asymmetry is corrected by 
$ A_{\mu} + A_{\text{track}} +A_{KK} = \left[ 0.13 \pm 0.06 \thinspace (\mbox{syst})\right]\%$, 
including the  statistical uncertainties combined in quadrature. The fraction of  \Bs\ 
decays in the sample and the time-integrated oscillation 
probability, we find  $F_{\Bs}^{\text{osc}} = 0.465$. The resulting time-integrated 
flavor-specific semileptonic charge asymmetry is found to be
\begin{align}
\asls = \left[ \rm{-1.12} \pm 0.74 \thinspace (\text{stat}) \pm 0.17 \thinspace (\text{syst}) \right]\%.
\end{align}

\section{The semi-leptonic charge asymmetry in \Bd\ mesons}

The measurement of the semi-leptonic charge asymmetry in \Bd\ mesons  is measured using two separate orthogonal decay channels:
\begin{enumerate}
\item $\Bd \rightarrow \mu^+ \nu D^{-} X$,  with $D^{-} \rightarrow K^+ \pi^- \pi^-$ (plus charge conjugate process);
\item $\Bd \rightarrow \mu^+ \nu D^{*-} X$, 
with $D^{*-} \rightarrow \bar{D}^0 \pi^-, \bar{D}^0 \rightarrow K^+ \pi^-$
(plus charge conjugate process);
\end{enumerate}
The two channels are treated separately, with each being used to extract $a^d_{\text{sl}}$, 
before the final measurements are combined.

The time-integrated  
flavor-specific semileptonic charge asymmetry in \Bd\ decays which have undergone flavor mixing, \asld ,  is then given by
 \begin{equation}
\asld \cdot {F_{\Bd}^{\rm{osc}}} = A - A_{\mu}  - A_{K} + 2 A_{\pi},
\label{Eq:asld}
\end{equation}
where $A_\pi = A_{\rm{track}}$ and $A_{K}$ is the residual kaon asymmetry.

The \Bd\ meson has a mixing frequency $\Delta M_d = 0.507 \pm 0.004$~ps$^{-1}$, 
of comparable scale to the lifetime $\tau(\Bd) = 1.518 \pm 0.007$~ps~\cite{pdg}. 
Hence the fraction of oscillated \Bd\ mesons is a strong function of the measured decay time. 
The best precision is obtained by binning the data as a function of the transverse decay length.  
Small values of the transverse decay length should have negligible contributions from oscillated \Bd\ mesons, and are 
not included in the final $a^d_{\rm{sl}}$ measurement. They represent a control region in which the 
measured raw asymmetry should be dominated by the background contribution.

\begin{figure*}[ht]
        \centering
        \subfigure[$\mu D$ channel]
        {\includegraphics[width=0.45\columnwidth]{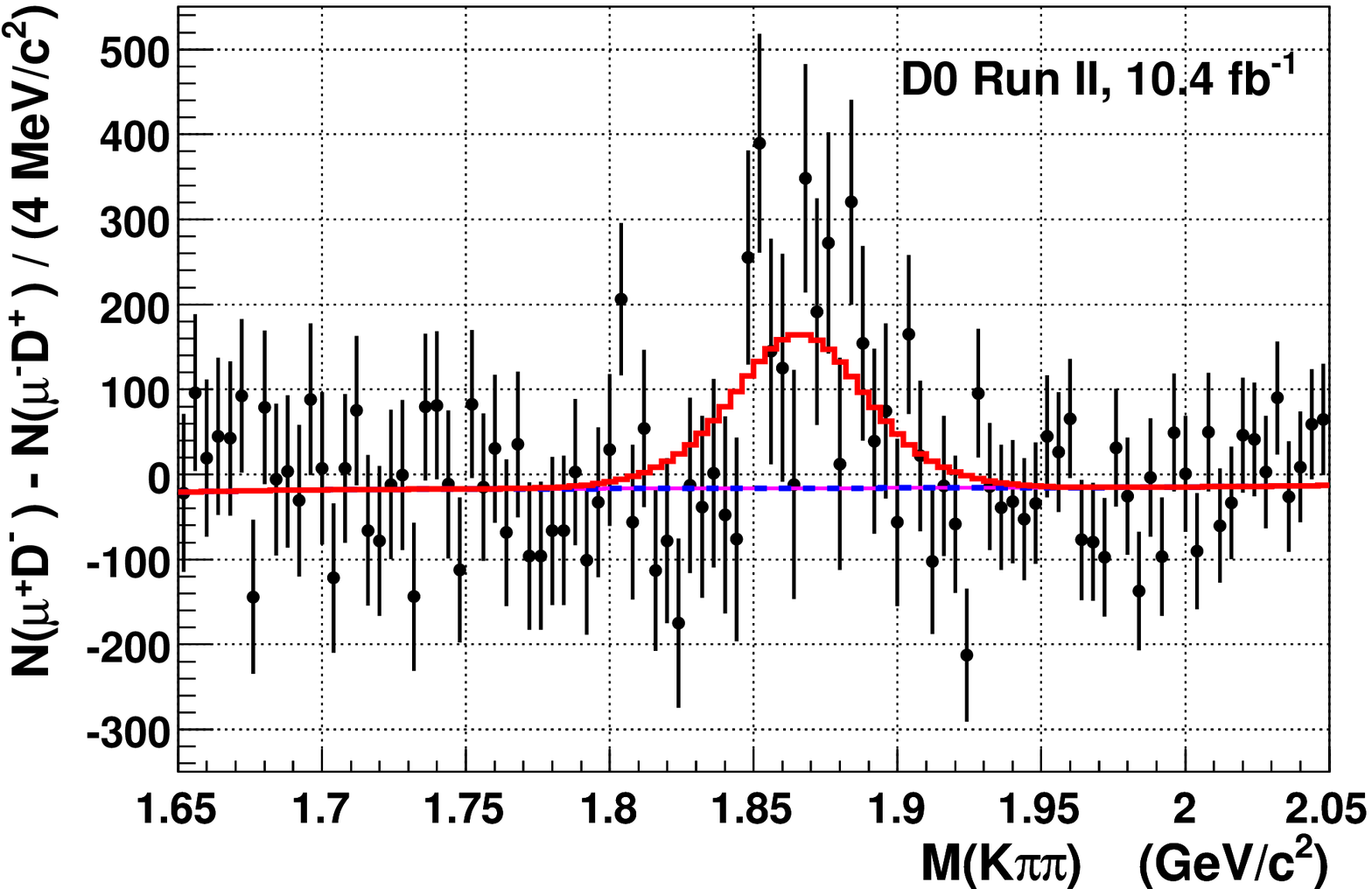}}
        \subfigure[$\mu D^*$ channel]
        {\includegraphics[width=0.45\columnwidth]{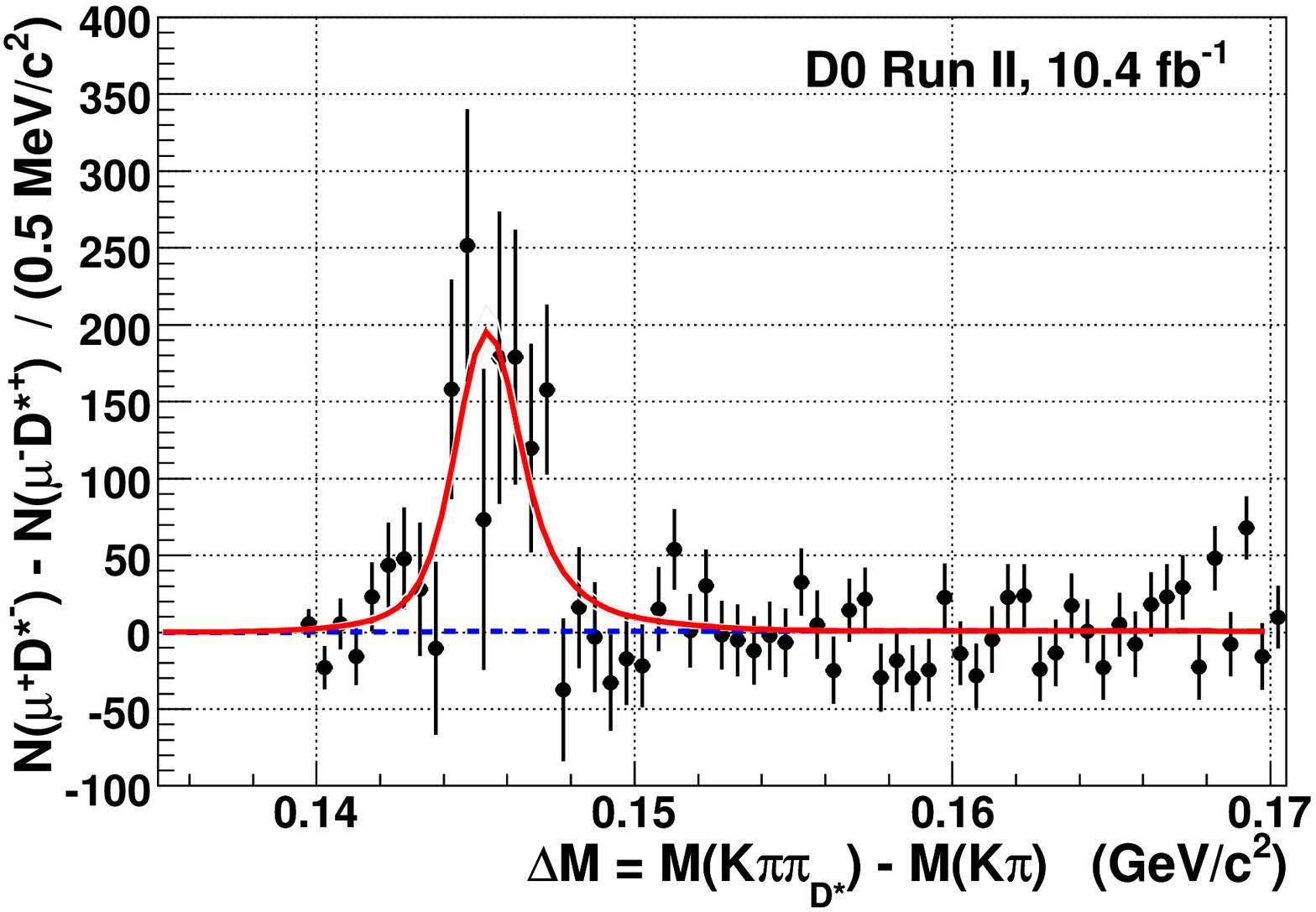}}
        \caption[]{Examples of the raw asymmetry fit for the two decays channels, 
          for the fifth visible proper decay length (\Bd ) bin corresponding to ($0.10 < \text{VPDL}(\Bd ) < 0.20$)~cm.
          The  plots show the difference distributions. 
          In both cases, the solid line represents the total fit function, with the background part 
          shown separately by the dashed line.}
\label{fig:rawresults}
\end{figure*}

The raw asymmetries for each decay channel and each decay length bin   are extracted using  $\chi^2$ 
minimization. 
The fits are performed simultaneously on the sum  and
the difference (Fig.~\ref{fig:rawresults}) of the $\mu D$ and $\mu D^\ast$ distributions.

The extracted raw asymmetries are corrected for background asymmetries on a bin-by-bin basis along with the systematic uncertainties.  Once the uncertainties on the individual $a^d_{\text{sl}}$ measurements are established, 
the combination between VPDL bins, and then between channels, is performed. 
For each channel, the combined $a^d_{\text{sl}}$ value is obtained by a weighted average 
of the four individual measurements where the weights are the inverse of the sum in quadrature of statistical 
and systematic uncertainties for that measurement.
The central values and uncertainties for the combinations are again determined by performing 
ensemble tests, with all inputs varied, and examining the effect on the 
final values of $a^d_{\rm{sl}}$ from each channel. This procedure yields the following results:
\begin{eqnarray}
a^d_{\text{sl}}(\mu D)   & = & [0.43 \pm 0.63 \text{ (stat.)} \pm 0.16 \text{ (syst.)}]\%,\text{ ~  ~} \\
a^d_{\text{sl}}(\mu D^*) & = & [0.92 \pm 0.62 \text{ (stat.)} \pm 0.16 \text{ (syst.)}]\%.\text{ ~  ~}
\end{eqnarray}
Finally, the combination is extended to give the full weighted average of the two channel-specific
measurements, with full propagation of uncertainties, to yield the final measurement:
\begin{eqnarray}
a^d_{\text{sl}} & = & [0.68 \pm 0.45 \text{ (stat.)} \pm 0.14 \text{ (syst.)}]\%.
\end{eqnarray}

\section{Combination of D0 results}

\begin{figure}[ht]
\centering
\includegraphics[width=0.8\columnwidth]{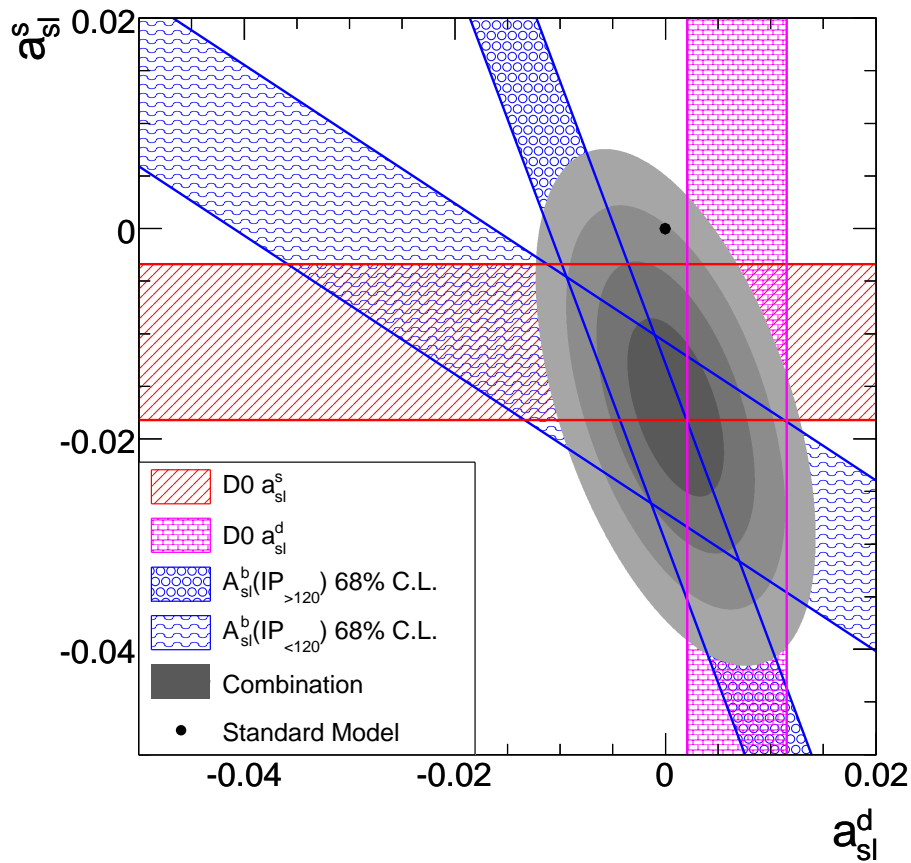}
\caption[]{Combination of measurements of the D0 measurements of $a^d_{\text{sl}}$, $a^s_{\text{sl}}$, and the two impact-parameter-binned constraints from the same-charge dimuon asymmetry $A^b_{\text{sl}}$ (D0~\cite{dimuon2}). The bands represent the $\pm 1$ standard deviation uncertainties on each measurement. The ellipses represent the 1, 2, 3, and 4 standard deviation two-dimensional confidence level regions of the combination.}
\label{fig:combo}
\end{figure}

The D0 measurements of $a^d_{\text{sl}}$, $a^s_{\text{sl}}$, can  be combined with the two-dimensional 
constraints on ($a^d_{\text{sl}}$, $a^s_{\text{sl}}$) from the D0 
measurement of the dimuon charge asymmetry $A^b_{\text{sl}}$~\cite{dimuon2}.
The full two-dimensional fit yields the following values:
\begin{eqnarray}
a^d_{\text{sl}}(\text{D0 comb}) & = & (0.10 \pm 0.30)\%, \\
a^s_{\text{sl}}(\text{D0 comb}) & = & (-1.70 \pm 0.56)\%, 
\end{eqnarray}
with a correlation coefficient of $-0.50$. The $\chi^2$ of this fit is 2.9, and the standard model $p$-value is $0.0036$,
corresponding to a 2.9 standard deviation effect. Figure~\ref{fig:combo} shows the two-dimensional contours from this combination.

\section{Updated World Averages}

\begin{figure}[ht]
\centering
\includegraphics[width=1.0\columnwidth]{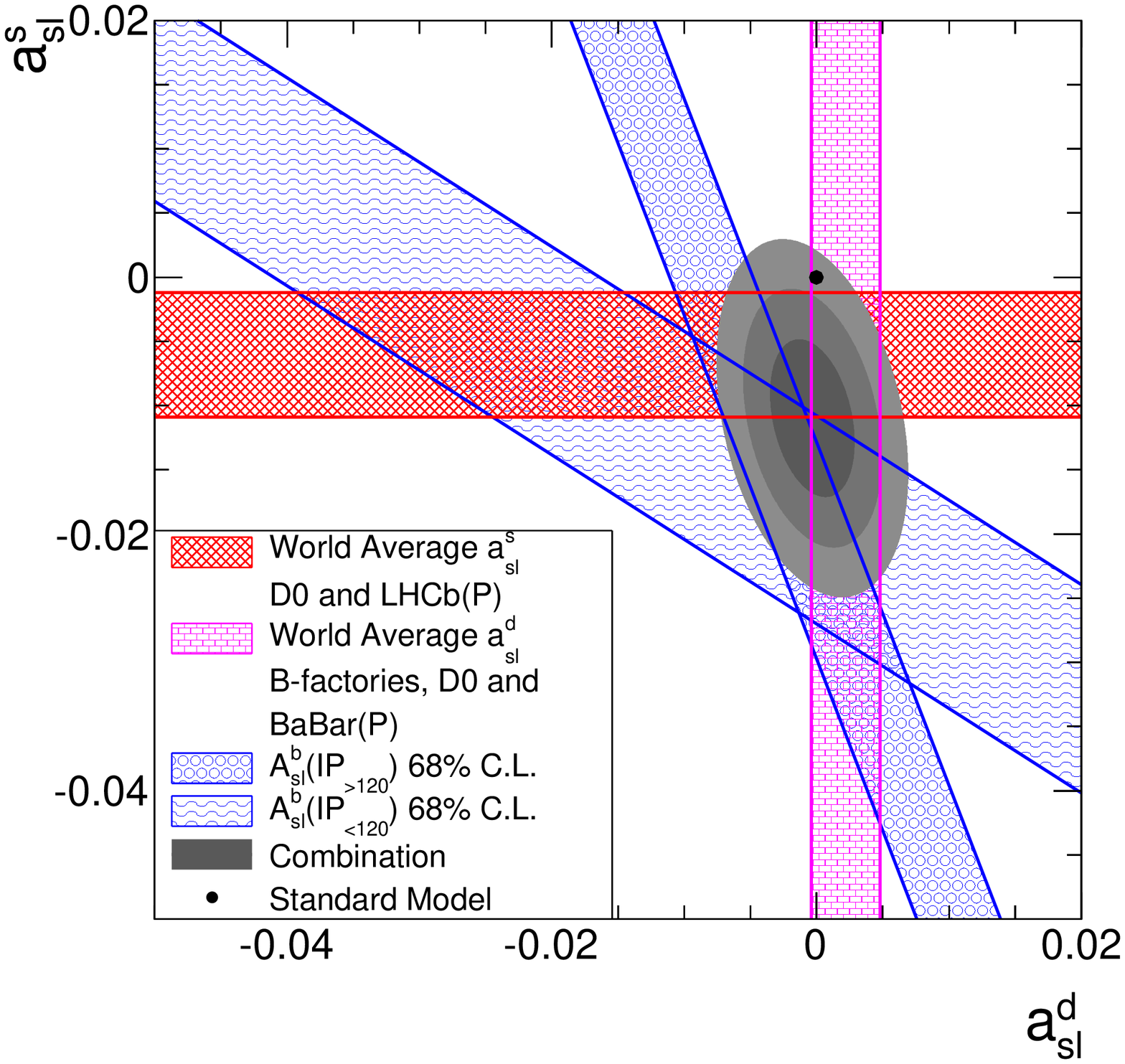}
\caption[]{Combination of measurements of the new world averages  of $a^s_{\text{sl}}$ (D0~\cite{d0asls} and LHCb~\cite{LHCb}), $a^d_{\text{sl}}$ 
($B$ factories average~\cite{pdg}, D0~\cite{d0adsl} and the new BaBar result~\cite{BaBar}),  and the two impact-parameter-binned constraints from the same-charge dimuon asymmetry $A^b_{\text{sl}}$ (D0~\cite{dimuon2}). The bands represent the $\pm 1$ standard deviation uncertainties on each measurement. The ellipses represent the 1, 2, and 3 standard deviation two-dimensional confidence level regions of the combination.}
\label{fig:comboAll}
\end{figure}

The D0 measurements of \asls\ and \asld\ can be combined with all other measurements to form updated world averages. I use a simple weighted average, assuming that the  measurements are fully independent. The D0~\cite{d0asls} and the preliminary LHCb~\cite{LHCb} measurements of \asls\ are combined: 
\begin{equation}
\asls (\rm{WA}) = \left( -0.60 \pm 0.49 \right) \%.
\end{equation}
The B-factories average~\cite{hfag}, the D0~\cite{d0adsl} and the new preliminary  BaBar~\cite{BaBar} measurement of \asld\ are also combined, 
\begin{equation}
\asld (\rm{WA}) = \left( 0.23 \pm 0.26 \right) \%.
\end{equation}
These numbers can  be combined with 
$A^b_{\text{sl}}$, we obtain the 
following values:
\begin{eqnarray}
a^d_{\rm{sl}}(\text{comb}) & = & (-0.03 \pm 0.21)\%, \\
a^s_{\rm{sl}}(\text{comb}) & = & (-1.10 \pm 0.41)\%, 
\end{eqnarray}
where the two parameters have a correlation coefficient of $-0.31$. The results are shown in Fig.~\ref{fig:comboAll}, with the 
two dimensional contours overlaid on the four constraints from the input measurements. 
The fit returns a $\chi^2$ of 3.8 for 2 degrees-of-freedom.
The $p$-value of the combination with respect to the SM point is $0.0148$, 
corresponding to an inconsistency at the 2.4 standard deviation level.

\begin{thebibliography}{99}
\bibitem{smprediction}
  A. Lenz and U. Nierste,  arXiv:1102.4274;
 A. Lenz and U. Nierste, J. High Energy Phys.  {\bf 06}, 072 (2007). 
 
 \bibitem{dimuon1} 
V.~M. Abazov {\it et al.} (D0 Collaboration), Phys. Rev. D {\bf 82}, 032001 (2010);  
V.~M. Abazov {\it et al.} (D0 Collaboration),   Phys. Rev. Lett. {\bf 105}, 081801 (2010). 

\bibitem{dimuon2} 
V. M. Abazov {\it et al.} (D0 Collaboration), Phys. Rev. D {\bf 84}, 052007 (2011). 

\bibitem {d0asls} 
V.~M.~Abazov \emph{et al.} (D0 Collaboration),   Phys. Rev. Lett., {\bf 110}, 011801 (2013).


\bibitem {d0adsl} 
V.~M.~Abazov \emph{et al.} (D0 Collaboration),  Phys.\ Rev.\  D. {\bf 86}, 072009 (2012).

   \bibitem{d0det}
V.M.~Abazov  \emph{et al.} (D0 Collaboration), Nucl. Instrum. Methods Phys. Res. A {\bf 565}, 463  (2006).


\bibitem{LHCb} LHCb Collaboration, LHCb-CONF-2012-022 (2012, unpublished). 

\bibitem{BaBar} C.~Cartaro, \emph{CP and T violation at BaBar and Belle} in the proceedings of  \emph{14th International Conference on B-Physics at Hadron Machines},
\pos{PoS(Beauty 2013)006}


\bibitem{pdg}
  J.~Beringer {\it et al.} (Particle Data Group), Phys.\ Rev.\ D {\bf 86}, 010001 (2012).


\end{thebibliography}
\end{document}